\documentclass{webofc}

\usepackage[varg]{txfonts}   
\usepackage{hyperref}
\usepackage{url}
\hypersetup{colorlinks=true,citecolor=blue,urlcolor=blue,linkcolor=blue}
\begin{document}
\title{3D Structure of Jet-induced Diffusion Wake}
%
%

\author{\firstname{Zhong} \lastname{Yang}\inst{1,2}\fnsep\thanks{\email{
    zhong.yang@vanderbilt.edu}}   \and
    \firstname{Yuxin} \lastname{Xiao}\inst{1}\fnsep\thanks{\email{
    xiaoyx@mails.ccnu.edu.cn}}  \and
    \firstname{Hanzhong} \lastname{Zhang}\inst{1}\fnsep\thanks{\email{
    zhanghz@ccnu.edu.cn}}  \and
        \firstname{Xin-Nian} \lastname{Wang}\inst{1}\fnsep\thanks{\email{xnwang@ccnu.edu.cn}}
}

\institute{Key Laboratory of Quark and Lepton Physics (MOE) \& Institute of Particle Physics, Central China Normal University, Wuhan 430079, China 
\and
           Department of Physics and Astronomy, Vanderbilt University, Nashville, TN
          }

\abstract{Jet-induced  diffusion wake in heavy-ion collisions  has a unique 3D structure as manifested in the jet-hadron correlations in rapidity and azimuthal angle. The rapidity asymmetry observable in dijets provides a robust measurement of the diffusion wake because it essentially background-free.
}
\maketitle
\section{Introduction}
\label{sec-1}

Jet quenching has been observed since around 2001 when experiments at RHIC  presented their first results. It was most clearly illustrated later by the first results on  dijet asymmetry from ATLAS at LHC~\cite{ATLAS:2010isq}. The increased dijet asymmetry in central Pb+Pb as compared to p+p collisions indicated strong jet quenching and energy loss. But important questions remain: where does the lost energy go, and how is it redistributed as jets propagate through the quark-gluon plasma (QGP)? Shortly after these first measurements and motivated by early two-particle correlation measurements by PHENIX \cite{PHENIX:2005zfm} , several proposals suggested that the energy deposited into the medium can act like a source of perturbation in a hot medium, and that these perturbations propagate like sound waves. Since jets propagate at the speed of light, which is much faster than the speed of sound in the medium, the perturbation will produce  Mach-cone like medium response. Hydrodynamic simulations also verified this picture: a shock wake front is followed by a diffusion wake behind the jet. The diffusion wake is essentially a depletion of energy or particle density in the medium caused by back-reaction processes of jet-medium interaction.

It was realized later on that the dihadron azimuthal correlation observed by PHENIX is caused by  triangular flow  due to event-by-event initial nuclear geometrical fluctuates.  More realistic numerical simulations of hadron correlation from jet-induced Mach-cone and experimental measurements of soft hadrons  associated with jets do not show any distinctive features that can be considered as ambiguous signals of the Mach-cone.  This is why we have to turn our attention to the signal of the diffusion wake.

\section{Effects of jet-induced medium response}
\label{sec-2}

We have used the coupled linear Boltzmann transport and hydro model (CoLBT-hydro)~\cite{Chen:2017zte} to investigate the effect of jet-induced Mach-cone excitation and the diffusion wake.  The LBT model~\cite{He:2015pra,Luo:2023nsi} uses the Boltzmann equation linearized around the medium background and considers only interactions between the jet shower and the medium partons, neglecting nonlinear interactions among medium excitations. Both $2\rightarrow 2$ and $2\rightarrow3$ processes are simulated according to pQCD and induced gluon radiation with LPM interference  is considered via the high-twist approach~\cite{Wang:2001ifa}. Crucially, recoil partons are allowed to rescatter, just like any other jet partons, leading to the propagation of medium response. Because jets are inherently 3D objects, we also need a full 3D hydrodynamics to properly simulate medium response. Assuming pQCD won't work for soft partons,  we consider partons below a cutoff (and the negative partons from the back-reaction) thermalize locally and contribute to a source term in the hydrodynamic evolution of the bulk medium. The hydrodynamics with the source term is used for the soft modes while hard partons evolve according to the LBT within the dynamically updated medium. That is the basic  idea behind the CoLBT approach.

\begin{figure}[h]
\centering
\includegraphics[width=4.cm,clip]{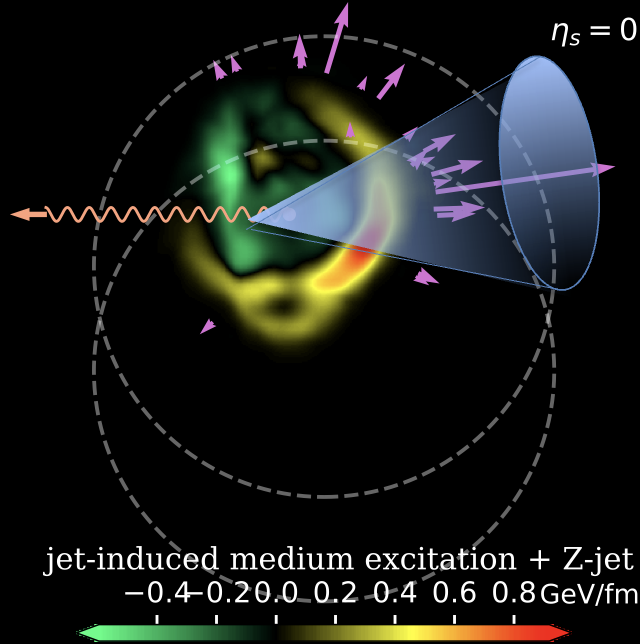}\hspace{0.5in}
\includegraphics[width=4.cm,clip]{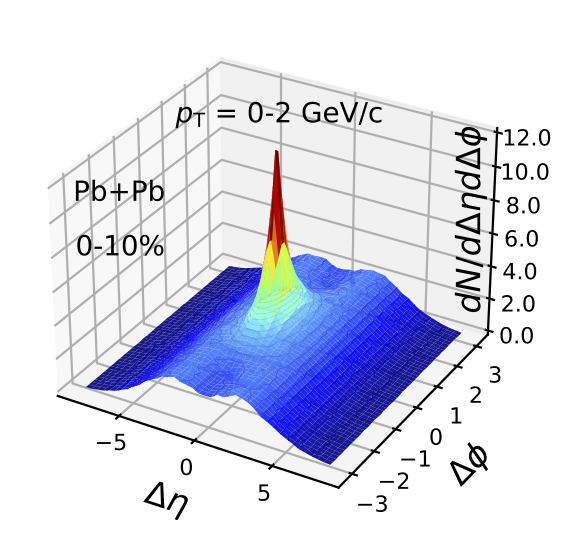} 
\caption{ (left) The energy density of the Z-jet induced medium response and jet partons (arrowed lines) in a non-central Pb+Pb collisions. (right) Jet-hadron correlation in rapidity and azimuthal angle in $\gamma$-jets in central Pb+Pb collisions at $\sqrt{s}$=5.02 GeV (figure from Ref.~\cite{Yang:2022nei}).}
\label{fig-0}       
\end{figure}

Shown in  Fig.~\ref{fig-0} (left) is an example of a Z+jet in a realistic non-central Pb+Pb collision event. In the recoil direction of the Z,  the jet deposits energy into the medium and creates a nonuniform energy density that varies event by event according to 3+1D viscous hydrodynamics. We run the hydrodynamic evolution with the same initial condition twice with and without the jet. The difference between the two hydrodynamic as shown in the figure is the medium response induced by the jet. One can clearly see the wake and the depleted region behind the jet. The wake shows a positive perturbation in the front (a shock/front) and a negative region behind, corresponding to the depletion. One can a clear angular correlations between the wake and energetic partons that are not fully quenched and escape.

One immediate consequence of the medium response is visible in the jet shape or the radial distribution of energy within a jet. Without recoil partons, one should see the usual suppression of leading hadrons and little enhancement at large radius. When medium response is included, an enhancement at large angle is seen that is dominated by soft hadrons. In the fragmentation function there is enhancement at small momentum fraction (soft hadrons) and suppression of the leading (hard) hadrons due to jet quenching. The large-angle soft enhancement, however, has two contributions: (1) medium response (the Mach-cone-like wake), and (2) hadronization of radiated gluons. In practice, it is very hard to separate these two contributions  experimentally.

\section{Signals of the diffusion wake}
\label{sec-3}

This is why we proposed to study the diffusion wake on the opposite side of the jet, since there is no analogous process by other mechanisms that produces a depletion of particles in the opposite direction of the jet. CMS has measured the $Z^0$-hadron angular correlations and, after subtracting underlying event contributions from multiple parton interaction (MPI), one can see a very small depletion on the away side of the jet ($Z^0$ side)~\cite{Yang:2021qtl}. The effect is, however, tiny and very hard to see in experimental data.

\begin{figure}[h]
\centering
\includegraphics[width=4.cm,clip]{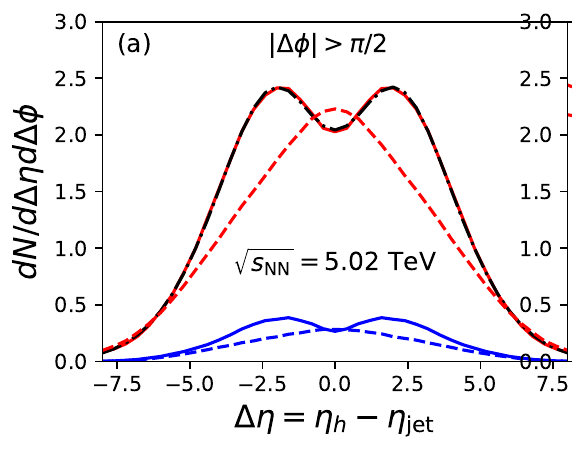} \hspace{0.5in}
\includegraphics[width=4cm,clip]{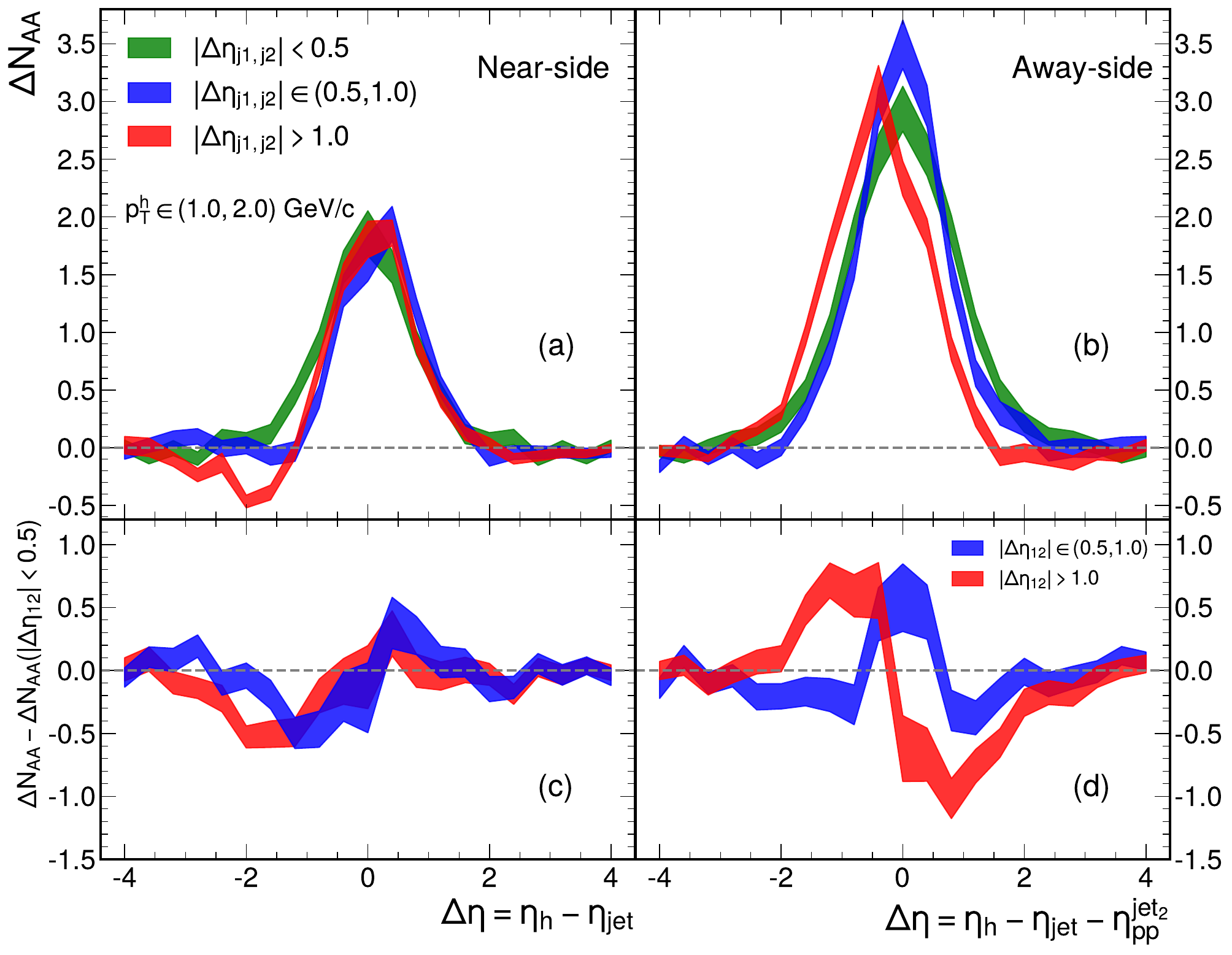}
\caption{ (left) Jet-hadron correlation in rapidity in the direction of the $\gamma$ in $\gamma$-jets for two different hadron pT ranges in central Pb+Pb collisions (solid) compared to p+p collisions (dashed) at $\sqrt{s}$=5.02 GeV (figure from Ref.~\cite{Yang:2022nei}).  (right) Jet-hadron correlations in rapidity (upper) and rapidity asymmetry in dijets in central Pb+Pb collisions at $\sqrt{s}=5.02$ GeV (figure from Ref.~\cite{Yang:2025dqu}). }
\label{fig-1}       
\end{figure}

We realized later on another important point:  Jets are 3D objects and so should be the diffusion wake. Besides the azimuthal correlations, one should examine the rapidity correlations as well. In our CoLBT calculations~\cite{Yang:2022nei}, we looked at the jet–hadron correlations in both rapidity and azimuthal angle. For $\gamma$+jet events in p+p collisions, the 2D correlation shows a peak in the jet direction with an underlying MPI contributions that are uniform in azimuthal angle but have a structure in rapidity similar to the underlying soft hadron production. We call this the MPI ridge. MPI is essentially independent mini-jet production in coincident with the hard scattering of $\gamma$-jet production. In central Pb+Pb collisions, soft hadrons from medium response and induced gluon-radiation enhance the jet peak, while a valley develops on top of the MPI ridge in the $\gamma$ direction due to the diffusion wake which  is very clear as shown in Fig.~\ref{fig-0} (right). When projecting onto the rapidity, one can observe a double-bump structure as shown in Fig.~\ref{fig-1} (left). The  big bump is from MPI and the dip is from the diffusion wake. Simulated results and analyses by other studies find similar and unique rapidity structures. We interpret this as an unambiguous signal of the diffusion wake. This structure of a dip on the MPI ridge has recently been observed by both ATLAS and CMS experiments~\cite{ATLAS:2024prm,CMS:2025dua}, providing the first direct observation of the diffusion wake and the jet-induced medium response.

Dijet events are trickier. The diffusion wake of one jet can be overwhelmed by the hard fragments of the opposing jet, making the effect hard to see. If two jets are separated in rapidity, the diffusion wake from one jet affects a different rapidity region from the other jet, so one can see a characteristic rapidity asymmetry. By subtracting events with and without a rapidity gap, the background largely cancels and the remaining difference is dominated by the diffusion wake~\cite{Yang:2025dqu}. This rapidity asymmetry observable as shown in Fig.~\ref{fig-1} (right) is more robust because it does not require explicit background subtraction.

We can also define a differential jet shape where we examine not just the radial distribution but the joint distribution in radius and azimuthal angle around the jet axis \cite{Xiao:2024ffk}. For events selected with given angle relative to the event-plane using gradient tomography, one should observe an azimuthal asymmetric jet shape. High-$p_T$ particles in the jet are deflected due to gradient of the medium; this deflection tends to be in the opposite direction of soft particles from the diffusion wake. Thus, in an azimuthal asymmetry distribution of the jet shape, you see a correlation between high-$p_T$   jet fragments and the soft particles from the wake. If the jet sits at larger rapidity, this feature shifts to a larger angle due to longitudinal flow coupling. One can therefore see the combined effects of the diffusion wake and longitudinal flow.

\begin{figure}[h]
\centering
\includegraphics[width=3.5cm,clip]{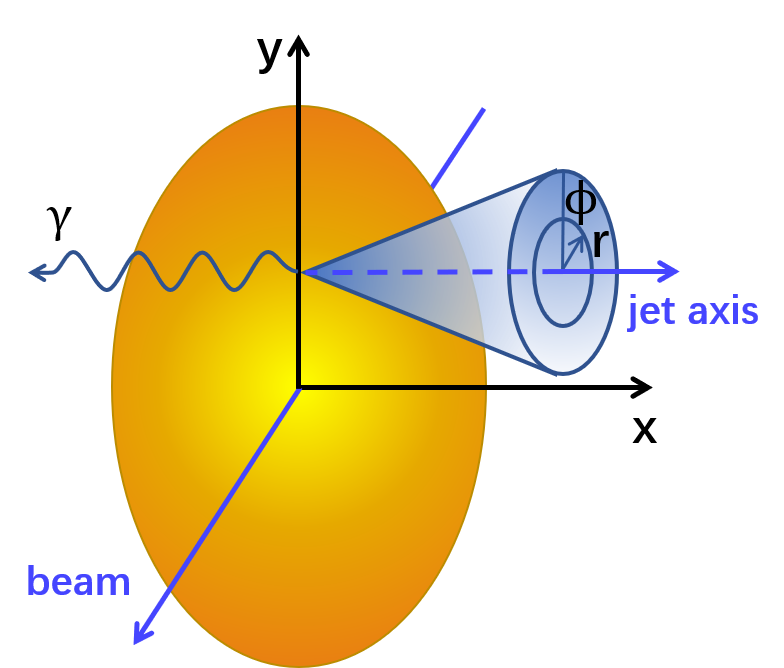} \hspace{0.5in}
\includegraphics[width=3.5cm,clip]{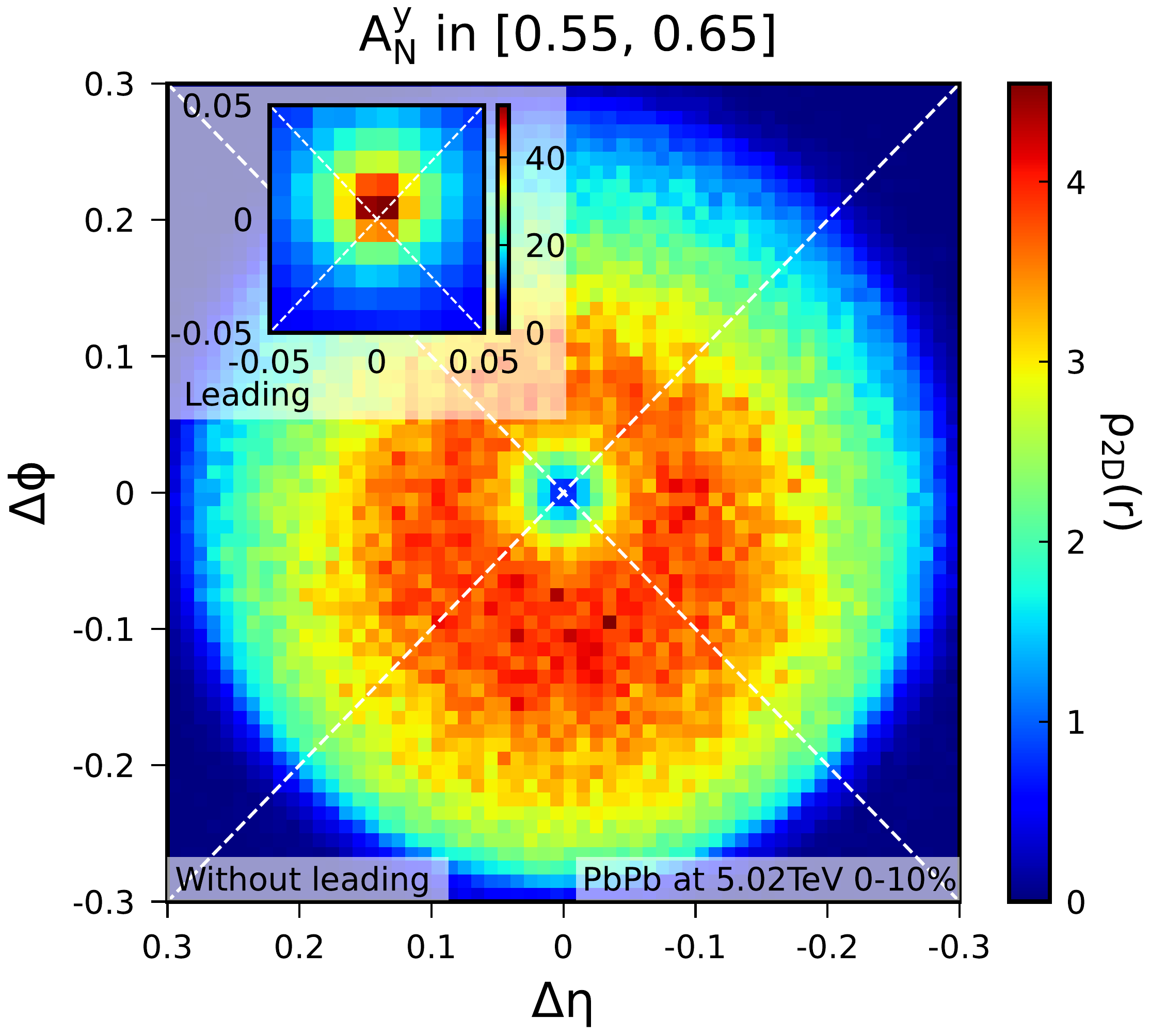}
\caption{ (left) Illustration of the azimuthal angle-dependent jet shape. (right) Azimuthal asymmetrical jet shape in central Pb+Pb collisions at $\sqrt{s}=5.02$ GeV (figure from Ref.~\cite{Yang:2025dqu}). }
\label{fig-2}       
\end{figure}


\section{Summary and Acknowledgements}
\label{sec-4}

In summary, medium response due to jet-medium interaction leads to an enhancement of soft hadrons in the jet direction and a depletion of soft hadrons in the opposite (back) direction due to the diffusion wake. The diffusion wake has a unique 3D structure leading to a unique feature in jet-hadron correlations in rapidity and azimuthal angle. The rapidity asymmetry observable provides a robust measurement of the diffusion wake because it essentially background-free. 

This work is supported in part by the China Postdoctoral Science Foundation under Grant No. 2024M751059 and No. BX20240134(ZY), by NSFC under Grant No. 12535010 and by the Guangdong MPBAR with No. 2020B0301030008.

%
%
%

\end{document}